\documentclass{phbauth}
\usepackage{epsf}
\usepackage{rotate}
\usepackage{graphicx}

\begin{document}
\begin{frontmatter}
\title{Transport in the Heavy Fermion Superconductor UPt$_3$}
\author[address1]{M. J. Graf\thanksref{thank1}},
\author[address2]{S.-K. Yip},
\author[address3]{J. A. Sauls}
\address[address1]{Center for Materials Science, 
Los Alamos National Laboratory, Los Alamos, New Mexico 87545, United States}
\address[address2]{Physics Division, 
National Center for Theoretical Sciences, Hsinchu 300, Taiwan}
\address[address3]{Department of Physics \& Astronomy, 
Northwestern University, Evanston, Illinois 60208, United States}
\thanks[thank1]{Corresponding author: email: graf@lanl.gov}

\begin{abstract}
We report new theoretical results and analysis for the transport properties of
superconducting UPt$_3$ based on the leading models for the pairing symmetry. 
We use Fermi surface data and the measured inelastic
scattering rate to show that the low-temperature thermal conductivity and 
transverse sound attenuation in the A and B phase of UPt$_3$ are in excellent
agreement with pairing states belonging to the two-dimensional orbital E$_{2u}$
representation.
\end{abstract}


\end{frontmatter}

  Much has been learned about the superconducting states
of the heavy fermion compound UPt$_3$ from studies of the 
superconducting phase diagram \cite{sauls94}. Transport properties have
played an important role in narrowing down the viable theoretical models
for the pairing state. Low-temperature measurements of the
thermal conductivity \cite{heat} have shown that the B-phase of 
UPt$_3$ is in quantitative
agreement with an order parameter (OP) belonging to either the 
$E_{1g}$ $E_{2u}$, or $AE$ pairing states,
but is not consistent with the models
based on accidentally degenerate representations belonging to the $AB$ 
classes or with the 1D orbital models based on triplet pairing and 
no spin-orbit coupling \cite{graf99a,norman96}.
  However, Tou {\it et al.} \cite{tou98} argue that measurements 
of the Knight shift in UPt$_3$ requires a non-unitary spin-triplet order
parameter with no spin-orbit locking \cite{machida96}. This interpretation 
conflicts with the observed anisotropic paramagnetic limiting
as well as the low-temperature transport measurements for the thermal 
conductivity \cite{choi91,graf99a}. Further analysis of the magnetic
and transport properties of the superconducting phases is required 
in order to resolve these apparent conflicts.
  
  Recently Ellman and co-workers \cite{ellman96} measured the
transverse ultrasonic absorption in both the A- and B-phases of
superconducting UPt$_3$ on the same crystals used to measure the
low-temperature thermal conductivity.
The results of our analysis show
that the pairing states belonging to the $E_{2u}$ representation are
quantitatively consistent with the thermodynamic phase
diagram, specific heat, anisotropic thermal conductivity and anisotropic
sound attenuation. Pairing states based on the $E_{1g}$ and $AE$
representations fail to account for the anisotropy in the sound
attenuation.

  We report the results of our calculations and analysis of the ultrasonic 
attenuation for both superconducting A and B phases in the 
long-wavelength hydrodynamic limit. The transverse viscosity
\mbox{\boldmath$\eta$} and the sound attenuation
\mbox{\boldmath$\alpha$} with wavevector $\bf q$ and polarization
\mbox{\boldmath$\varepsilon$} are related by
\begin{equation}
\alpha_{i j}(T) = {\omega^2}/({\varrho c_s^3}) \,
        \eta_{i j, k l}({\bf q}, \omega)\,
        \hat\varepsilon_i \hat q_j
        \hat\varepsilon_k \hat q_l \,,
\end{equation}
where $\varrho$ is the mass density, $c_s$ the speed of sound,
and $\omega = c_s q$, and \mbox{\boldmath$q \cdot \varepsilon $}$= 0$.
For transverse waves with polarization
\mbox{\boldmath$\varepsilon$}$ || \hat x$
propagating along ${\bf q} || \hat y$, or vice versa,
all relevant OP models
possess a {\it universal} zero-temperature value,
except for the $AE$-model.
In the case of the $E$-models the viscosity is
$\eta_{x y,x y}(0,0) \simeq v_f^2 p_f^2 N_f /
 | 8\, d\Delta(\theta)/d\theta|_{\rm node}$. For other
propagation directions the transverse viscosity has a nonuniversal but
finite value for $\omega, T \to 0$ \cite{graf96c}.

The anomaly in the attenuation at the A-B transition
(Fig. 1) is largest for $\alpha_{x y}$, and only weakly visible 
in $\alpha_{x z}$. The anomaly reflects a decrease in the number of thermally 
excited quasiparticles in the B-phase relative to 
the A-phase. As temperature decreases below the second superconducting
transition, $T_{c-}$, the subdominant OP 
nucleates and closes off the additional nodes in the A-phase. 
The anisotropy of this anomaly provides new information on the nodal 
structure of the order parameter.

\begin{figure}
  \noindent
  \begin{minipage}{0.48\textwidth}
  \begin{center}
    \mbox{
      \epsfysize=0.96\hsize
      \rotate[r]{\epsfbox{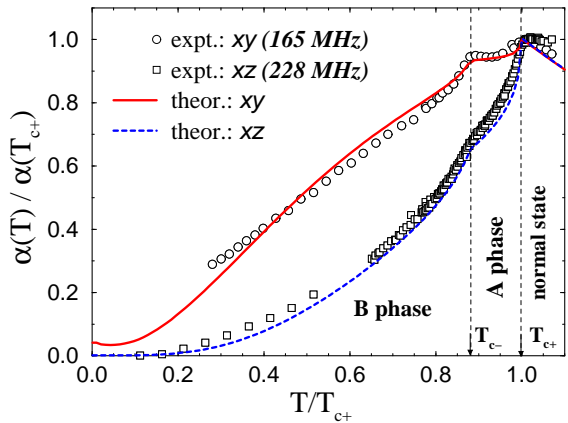}}
    }
    \caption[]{
    Calculations of the transverse
    sound attenuation for an $E_{2u}$ OP in the limit $\omega\to 0$ 
    with a phenomenological scattering rate
    $\Gamma(T) = 0.01\pi T_{c+} (1+T^2/T_{c+}^2)$.
    The splitting of $T_{c+}-T_{c-}$ in the specific heat determines 
    the bare transition temperatures $T_{c2}/T_{c1}=0.92$.
    The experimental data are from Ellman {\it et al.}~\cite{ellman96}.
    }\label{UPt3}
  \end{center}
  \end{minipage}
\end{figure}

  Our calculations show that the reported enhancement of the sound attenuation
in the A-phase is in excellent agreement with the $(1,0)$ state of 
an $E_{2u}$ OP model in the resonant impurity scattering limit;
see Fig.~\ref{UPt3}.  The order parameter
and the scattering parameters are exactly those obtained from
the analysis of the thermal conductivity on the same crystal~\cite{graf99a}.
Thus, there are no adjustable parameters in the calculation
of the sound attenuation. 
The  $E_{1g}$ and $AE$ models also provide an excellent fit to the thermal 
conductivity data, but fail to account for the transverse sound attenuation 
anomaly.  This difference reflects the differences in the nodal structure 
for the A-phase in the two different $E$-rep models and the $AE$ model.

  In conclusion, measurements of $\alpha_{x y}$ and $\alpha_{x z}$ are
in excellent agreement with the $(1,0)$ orbital state
in the high-temperature A-phase with $E_{2u}$ symmetry and a
$k_z (k_x^2 - k_y^2)$ nodal structure.  The low-temperature
B-phase is described by a $(1,i)$ orbital state.

{\it Acknowledgement:}\ 
This research was supported by the NSF (DMR 91-20000) through the STCS.
MJG also acknowledges support from the Los Alamos National
Laboratory under the auspices of the DOE.

\end{document}